\input harvmac.tex
%\draftmode

\let\includefigures=\iftrue
\newfam\black
\includefigures
\input epsf
\def\figin{\epsfcheck\figin}\def\figins{\epsfcheck\figins}
\def\epsfcheck{\ifx\epsfbox\UnDeFiNeD
\message{(NO epsf.tex, FIGURES WILL BE IGNORED)}
\gdef\figin##1{\vskip2in}\gdef\figins##1{\hskip.5in}% blank space instead
\else\message{(FIGURES WILL BE INCLUDED)}%
\gdef\figin##1{##1}\gdef\figins##1{##1}\fi}
\def\DefWarn#1{}
\def\figinsert{\goodbreak\midinsert}
\def\ifig#1#2#3{\DefWarn#1\xdef#1{fig.~\the\figno}
\writedef{#1\leftbracket fig.\noexpand~\the\figno}%
\figinsert\figin{\centerline{#3}}\medskip\centerline{\vbox{\baselineskip12pt
\advance\hsize by -1truein\noindent\footnotefont{\bf Fig.~\the\figno:} #2}}
\bigskip\endinsert\global\advance\figno by1}
%%%
\else
\def\ifig#1#2#3{\xdef#1{fig.~\the\figno}
\writedef{#1\leftbracket fig.\noexpand~\the\figno}%
%\figinsert\figin{\centerline{#3}}\medskip\centerline{\vbox{\baselineskip12pt
%\advance\hsize by -1truein\noindent\footnotefont{\bf Fig.~\the\figno:} #2}}
%\bigskip\endinsert
\global\advance\figno by1}
\fi

\Title{\vbox{\baselineskip12pt\hbox{hep-th/0003215}
\hbox{CALT-68-2267}\hbox{CITUSC/00-016}\hbox{SU-ITP-00-11}}}
{\vbox{ \centerline{Noncommutative
Gauge Dynamics From} \vskip 10pt \centerline{The String Worldsheet} }}
\centerline{Jaume Gomis$^1$, Matthew Kleban$^{2,3}$, Thomas Mehen$^1$,
Mukund Rangamani$^{1,4}$, Stephen Shenker$^2$} \medskip

\medskip \medskip \medskip \centerline{$^1$\it California Institute of
Technology, Pasadena, CA 91125 } \centerline{\it CIT-USC Center for
Theoretical Physics} \medskip \medskip \centerline{$^2$\it Department of
Physics, Stanford University, CA 94305-4060} \medskip \medskip
\centerline{$^3$\it Department of Physics, University of California,
Berkeley, CA 94720} \medskip \medskip \centerline{$^4$\it Department of
Physics, Princeton University, NJ 08544}

\medskip
\medskip
%\centerline{\tt }
\medskip
%\bigskip
\noindent
\medskip
\medskip

We show how string theory can be used to reproduce the one-loop
two-point photon amplitude in noncommutative $U(1)$ gauge theory.
Using a simple realization of the gauge theory in bosonic string
theory, we extract from a string cylinder computation in the
decoupling limit the exact one loop field theory result. The
result is obtained entirely from the region of moduli space where
massless open strings dominate. Our computation indicates that the
unusual IR/UV singularities of noncommutative field theory do not
come from closed string modes in any simple way.

\smallskip
%\medskip

\Date{March 2000}

\newsec{Introduction}
\nref\cds{A. Connes, M.R. Douglas and A. Schwarz, ``Noncommutative
Geometry and Matrix Theory: Compactification on Tori'', JHEP {\bf
9802}(1998) 003, hep-th/9711162.}%
\nref\dh{M.R. Douglas and C. Hull, "D-branes and the Noncommutative
Torus", JHEP {\bf 9802} (1998) 008, hep-th/9711165.}%
\nref\dhn{B. de Wit, J. Hoppe and H. Nicolai, ''On The Quantum
Mechanics of Supermembranes," Nucl. Phys. {\bf B305} (1988) 545.}
\nref\chekro{ Y.-K. E. Cheung and M. Krogh, ''Noncommutative Geometry From
0-Branes In A Background B Field," Nucl. Phys. {\bf B528} (1998)
185, hep-th/9803031}
\nref\chuho{C.-S. Chu and P.-M. Ho, "Noncommutative Open String
And D-Brane," Nucl. Phys. {\b550}(1999)151, hep-th/9812219;
''Constrained Quantization of Open String In Background B Field
and Noncommutative D-brane," hep-th/9906192.}
\nref\scho{V. Schomerus, ''D-Branes And Deformation Quantization,"
JHEP {\bf 9906:030} (1999), hep-th/9903205.}
\nref\aas{F. Ardalan, H. Arafei and M. M. Sheikh-Jabbari,
''Mixed Branes and M(atrix) Theory on Noncommutative Torus,"
hep-th/980367; "Noncommutative Geometry From Strings and Branes,"
JHEP {\bf 02} 016 (1999) hep-th/9810072; ''Dirac Quantization of
Open Strings and Noncommutativity in Branes," hep-th/9906161.}
\nref\bigsus{D. Bigatti and L. Susskind, hep-th/9908056 .}
\nref\seibwitt{N. Seiberg and E. Witten, `` String
Theory and Noncommutative Geometry'', JHEP {\bf 9909} (1999) 032,
hep-th/9908142.}%

It is striking that field theories on noncommutative spaces are
naturally embedded in string theory.  The first complete example
of this phenomenon was found in  toroidal compactifications of
Matrix Theory with
a nonzero $B$ field \cds \dh.   Of course the remarkable early
construction of membranes by large matrices \dhn\ is very much in this
spirit.  After \cds\dh\ additional work was done on extracting
noncommutative Yang-Mills theory more directly from open string
theory \chekro-\bigsus.   In a sense this work culminated in
\seibwitt\ where, among other things, a decoupling limit was
carefully formulated in which perturbative open string theory
reduced to noncommutative Yang-Mills theory.
%%%
\nref\filk{T. Filk, "Divergences in a Field Theory on Quantum Space",
Phys. Lett. {\bf B376} 53.}%
\nref\vgra{J.C. Varilly and J.M. Gracia-Bondia, "On the ultraviolet
behaviour of quantum fields over noncommutative manifolds",
Int. J. Mod. Phys. {\bf A14} (1999) 1305, hep-th/9804001.}%
\nref\two{M. Chaichian, A. Demichev and P. Presnajder, ``Quantum Field
Theory on Noncommutative Space-times and the Persistence of
Ultraviolet Divergences'', hep-th/9812180;  ``Quantum Field Theory on
the Noncommutative Plane with E(q)(2) Symmetry'', hep-th/9904132.}%
\nref\rruiz{C.P.Martin, D. Sanchez-Ruiz,
``The One-loop UV Divergent Structure of U(1) Yang-Mills
 Theory on Noncommutative $R^4$",
Phys. Rev. Lett. {\bf 83} (1999) 476-479,hep-th/9903077}%
\nref\rjabbari{ M. Sheikh-Jabbari, ``Super Yang-Mills Theory on
 Noncommutative Torus from Open String Interactions'',
 Phys. Lett. {\bf B450} (1999) 119, hep-th/9810179;
``One Loop Renormalizability
of Supersymmetric Yang-Mills Theories on Noncommutative Torus",
 JHEP {\bf 06} (1999) 015, hep-th/9903107; ``On the Deformation of
 $\Lambda$-Symmetry in B-field Background'', Phys. Lett. {\bf B477}
 (2000) 325, hep-th/9910258;
"Noncommutative Super
Yang-Mills Theories with 8 Supercharges and Brane Configurations",
hep-th/0001089.}%
\nref\gkw{H. Grosse, T. Krajewski and R. Wulkenhaar, "Perturbative
quantum gauge fields on the noncommutative torus", hep-th/9903187.}%
\nref\twopfo{S. Cho, R. Hinterding, J. Madore and H. Steinacker,
``Finite Field Theory on Noncommutative Geometries'',
hep-th/9903239.}%
\nref\three{E. Hawkins, ``Noncommutative Regularization for
Practical Man'', hep-th/9908052.}%
\nref\riouri{I. Chepelev and  R. Roiban,
``Renormalization of Quantum Field Theories on Noncommutative $R^d$, I.
Scalars,'' hep-th/9911098.}%
\nref\arcioni{G. Arcioni and M. A. Vazquez-Mozo,
``Thermal effects in perturbative noncommutative gauge theories'',
JHEP {\bf 0001} (2000) 028, hep-th/9912140.}%
\nref\ikk{S. Iso, H. Kawai and Y. Kitazawa,
``Bi-local fields in noncommutative field theory'', hep-th/0001027.}%
\nref\gkw{H. Grosse, T. Krajewski and R. Wulkenhaar, "Renormalization
of noncommutative Yang-Mills theories: A simple example", hep-th/0001182.}%
\nref\yaaa{I.Ya. Aref\'eva, D.M. Belov and A.S. Koshelev, "Two-Loop
Diagrams in Noncommutative $\phi^4_4$ theory", hep-th/9912075;
"A Note on UV/IR for Noncommutative Complex Scalar Field",
hep-th/0001215.}%
\nref\yaaa{I.Ya. Aref\'eva, D.M. Belov, A.S. Koshelev and
O.A. Rytchkov, "UV/IR Mixing for Noncommutative Complex Scalar Field
Theory, II (Interaction with Gauge Fields)", hep-th/0003176.}%
\nref\arsad{F. Ardalan and N. Sadooghi, "Axial Anomaly in
Non-Commutative QED on $R^4$",  hep-th/0002143.}%
\nref\grama{J.M. Gracia-Bondia and C.P. Martin, "Chiral Gauge
Anomalies on Noncommutative $R^4$", hep-th/0002171.}%
\nref\mvs{S. Minwalla, M.V. Raamsdonk and N. Seiberg, ``Noncommutative
Perturbative Dynamics'', hep-th/9912072.}%
\nref\vs{M.V. Raamsdonk and N. Seiberg, `` Comments on Noncommutative
Perturbative Dynamics'', hep-th/0002186.}%
\nref\haya{M. Hayakawa,``Perturbative analysis on infrared
and ultraviolet aspects of
noncommutative QED on $R^4$,'' hep-th/9912167.}%
\nref\texa{W. Fischler, E. Gorbatov, A. Kashani-Poor, S. Paban,
P. Pouliot and J. Gomis, ``Evidence for winding states in
noncommutative quantum field theory,'' hep-th/0002067.}%
\nref\sunew{A. Matusis, L. Susskind and N. Toumbas,
``The IR/UV connection in the non-commutative gauge theories,''
hep-th/0002075.}%
%%%
A large amount of work, partially inspired by these developments,
has been done on the perturbative dynamics of noncommutative field
theories \filk-\sunew.  These field theories
have a very interesting and unusual
perturbative behavior \mvs-\sunew.
The noncommutativity of the underlying space gives rise to a
strong mixing between the ultraviolet and the infrared \mvs\vs .
There are analogs of this IR/UV mixing in string theory which
provides one motivation to study these systems.
When loop
diagrams are evaluated in these theories, large momentum regions of the loop
integration lead to terms in the effective action that are infrared divergent
and nonanalytic in the noncommutativity parameter $\theta$.  In conventional
field theory, singularities in the low energy effective action usually
reflect omission of relevant low energy degrees of freedom, and the low
energy description is cured once the the missing degrees of freedom are
added. In \mvs\vs, it was proposed that at least some of
the novel IR/UV divergences of one
loop diagrams in the noncommutative theory could be understood as arising from
tree level exchange of new degrees of freedom. This phenomenon is
analogous to open-closed channel duality in one loop string graphs, where
ultraviolet divergences in the open string channel can be interpreted as
infrared divergences arising from tree level exchange of massless closed
strings. This work is motivated in part by trying to understand this
interpretation of the IR/UV singularities in noncommutative field
theories.

Noncommutative gauge theories can be realized in string theory by taking a
low energy decoupling limit of theories on D-branes, in the presence of a
constant magnetic field \cds\dh\seibwitt.  In this stringy setup one can try
to reproduce the noncommutative perturbative expansion from string theory.
This embedding confronts the issue of whether extra degrees of freedom --
apart from the obvious massless open string modes -- are needed to make sense
of the low energy effective action. As we will show, there seems to be no
need to add any further degrees of freedom.  We can account for the entire
field theory result by looking at the region of cylinder moduli space which
is dominated by massless open strings. The opposite region of moduli space,
where massless closed strings dominate, gives a vanishing contribution in the
zero-slope limit and therefore seem to decouple in the field theory limit.
This seems to indicate that the degrees of freedom proposed in \mvs\vs , even
though they reproduce the low energy effective action after integrating them
out, do not have a natural interpretation as massless closed string states.

In section $2$ we compute the two-point function of  the pure
noncommutative $U(1)$ gauge theory in 3+1 dimensions at one loop in the
background field gauge. The background field gauge is very useful for
comparing field theory amplitudes with the zero slope limit of string
amplitudes because the effective action obtained in the background field
gauge is manifestly gauge invariant, as is the answer obtained in string
theory.  At one loop, the two point function of the noncommutative gauge
theory has terms which contain logarithmic and quadratic \haya\sunew\
infrared divergences which do not appear in conventional gauge theories. The
appearance of quadratic infrared divergences is surprising, but nevertheless
compatible with gauge invariance.

In section $3$ we embed this gauge theory in bosonic string theory by
considering the low energy limit of the theory on a single D3-brane stuck
\nref\mrd{M.R. Douglas, "Enhanced Gauge Symmetry in M(atrix) Theory",
JHEP {\bf 9707} (1997) 004, hep-th/9612126.}%
\nref\ddg{D.E. Diaconescu, M.R. Douglas and J. Gomis, "Fractional
Branes and Wrapped Branes", JHEP {\bf 9802} (1998) 013,
hep-th/9712230.}%
at an $R^{22}/Z_2$ orbifold singularity and  in the presence of a magnetic
field along the worldvolume directions \mrd\ddg \foot{In this paper we do not
discuss the physics of noncommutative field theories with $\theta^{0i}\neq
0$. Such a gauge theory can be realized by having a constant electric field
along the brane. However, in the decoupling limit the upper bound of the
allowed electric field vanishes and the string theory realization is
ill-defined.}.
 We compute at one loop the planar
and non-planar contributions to the two-point function of photons in the
string theory. The field theory answer is reproduced by isolating from the
string theory amplitude the contribution coming from the boundary of cylinder
moduli space where massless open strings are important. Our result indicates
that closed string modes decouple from the low energy noncommutative field
theory, just as they decouple from conventional gauge
theories realized on branes. We conclude with a discussion of our
results in section $4$.

\newsec{Gauge Theory Calculation}

In this section we describe the calculation of the two point function
 in
 noncommutative
 $U(1)$ gauge theory in the background field gauge.  The action for
 the U(1) noncommutative Yang-Mills theory in a background metric $G^{\mu\nu}$
is given by
\eqn\action{ S = - {1 \over 4} \int d^4x\sqrt{-G}G^{\mu\rho}G^{\nu\sigma}
F_{\mu \nu}\star F_{\rho
 \sigma},}
where the field strength is \eqn\fieldstr{\eqalign{ F_{\mu \nu} =
\partial_{\mu} A_{\nu} -  \partial_{\nu} A_{\mu} - i g [A_{\mu}, A_{\nu} ]
\cr [ A_\mu, A_\nu ] = A_{\mu} \star A_{\nu} - A_{\nu} \star A_{\mu}.}} The
action in \action\ is invariant under the gauge transformation \eqn\gauge{
\delta_{\lambda} A_{\mu} = \partial_{\mu} \lambda- i g [A_\mu,  \lambda]
\equiv D_\mu \lambda.} The noncommutative star product appearing in
\action\fieldstr\ is defined by \eqn\starproduct{ f(x) \star g(x) = e^{{i
\over 2} \theta^{i j} {\partial \over \partial \alpha_i}{\partial \over
\partial \beta_j}} f(x+\alpha) \, g(x+\beta)|_{\alpha=\beta=0} , } where the
parameter $\theta^{i j}$ is related to the commutator of the coordinates in
the noncommutative space: \eqn\com{ [x^\mu,x^\nu] = i \theta^{\mu \nu} .}

We will quantize the theory in background field gauge
\nref\bgf{ B.S. DeWitt, Phys. Rev. 162 (1967) 1195, 1239; {\it in}
Dynamic Theory of Groups and Fields (Gordon and Breach, 1965)}%
\nref\abb{L. F. Abbot, Nucl. Phys. {\bf B185} (1981) 189-203.}%
\bgf\abb. The gauge field $A_\mu$ is split into classical and a quantum
pieces denoted $A_\mu$ and $Q_\mu$  respectively. The path integral is
performed over the quantum fields while the classical fields are kept fixed.
The generating functional for Green's functions is given by
\eqn\Z{ Z[J,A] =\hskip-5pt
\int [dQ] {\rm det}[{\delta \Delta\over \delta \lambda}] {\rm Exp}
\left[\hskip-2pt  i
\int d^4x{\sqrt -G}
 \left(  - {1 \over 4} G^{\mu\rho}G^{\nu\sigma} F_{\mu \nu}\star F_{\rho
 \sigma}
-{1 \over 2 \alpha} \Delta^2 +  G^{\mu\nu}J_\mu Q_\nu\right) \right],}
where $\Delta$ is the gauge
fixing condition, ${\rm det}[{\delta \Delta/ \delta \lambda}]$ is the
Faddeev-Popov determinant and $J_\mu$ is an external current coupled only to
the quantum fields. In \Z , $F_{\mu \nu}$ is understood to be a function of
both $A_\mu$ and $Q_\mu$. The background field gauge effective action is
defined by \eqn\effaction{ \Gamma[{\bar Q},A] = W[J,A] -\int d^4x {\sqrt -G}
G^{\mu\nu}J_\mu {\bar Q}_\nu ,} where \eqn\defWQ{ W[J,A] = -i \,{\rm ln}\,
Z[J, A]
\,\,\,\,\,\, {\rm and} \,\,\,\,\,\,{\bar Q}_\mu = {\delta W \over \delta
J^\mu} . } The background field gauge effective action is invariant under the
transformations \eqn\trans{\eqalign{ \delta A_\mu &= \partial_\mu \lambda - i
g [A_\mu, \lambda] \cr \delta {\bar Q}_\mu &= [\lambda,{\bar Q}_\mu], }} so
$\Gamma[0, A]$ is a manifestly gauge invariant functional of the classical
field $A_\mu$.
\ifig\ncfr{Feynman rules for Noncommutative $U(1)$ in background field
gauge. In the calculation of section 2, we use Feynman gauge, $\alpha =1$.}
{\epsfxsize6in\epsfbox{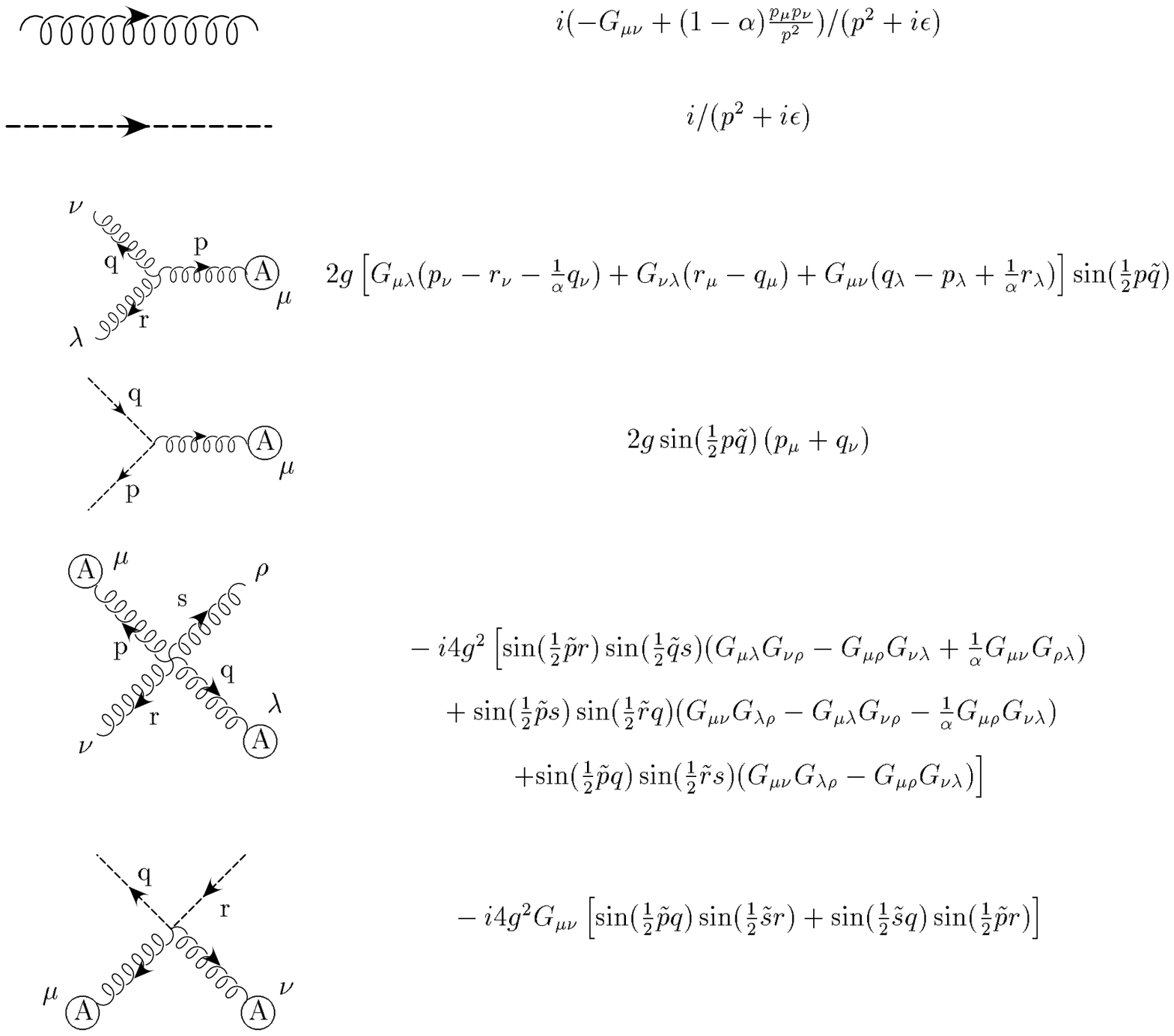}}
We can compute $\Gamma[0, A]$ by summing one-particle
irreducible Feynman diagrams with classical fields $A_\mu$ on external legs
and quantum fields $Q_\mu$ appearing only in internal lines. The Feynman
rules for conventional non-Abelian gauge theory in the background field gauge
 are given in \abb . The Feynman rules for
the noncommutative $U(1)$ theory can be obtained from \abb\ by simply
replacing the structure constants $f_{abc}$ by ${\rm sin}({1 \over 2} p_\mu
\theta^{\mu \nu} k_\nu)$, where $p,k$ are the momenta of two of the gluons
entering the vertex. The Feynman rules relevant for the calculations of this
paper are shown in \ncfr .

The explicit gauge invariance of $\Gamma[0,A]$ simplifies computations in the
gauge theory. The bare field strength, when expressed in terms of the
renormalized coupling and gauge field, is \eqn\renfs{ F_{\mu \nu}^{bare} =
Z_A^{1/2}(\partial_{\mu} A_{\nu} -
 \partial_{\nu}
 A_{\mu} - i g Z_g Z_A^{1/2}
[A_{\mu}, A_{\nu} ] ) ,}
where the $Z_g, Z_A$ are the coupling constant and field renormalizations
\eqn\ZAZg{
g^{bare} = Z_g g , \,\,\,\, A^{bare}_\mu = Z_A^{1/2} A^{bare}_\mu .}
Gauge invariance of \renfs\ implies that $Z_A^{-1/2} =Z_g$. This means
that the $\beta$-function can be computed from $Z_A$, which only
requires knowledge of the two-point function of the theory.

\nref\mettesy{R.R. Metsaev and A.A. Tseytlin, ``One Loop Corrections
to String Theory Effective Actions'', Nucl. Phys. {\bf B298} 109 (1988).}%
The background field gauge is also very useful for comparing field theory
amplitudes with the zero slope limit of string theory amplitudes.
Ref. \mettesy\  derived the $\beta$-function of Yang-Mills theory by
calculating the effective action of strings in a background magnetic field.
\nref\bd{Z. Bern and D. C. Dunbar, Nucl. Phys. {\bf B379}(1992) 567-601.}%
Ref. \bd\ pointed out a correspondence between loop amplitudes in
the background field gauge and loops calculated using string motivated rules.
\nref\DiVec {P. Di Vecchia, A. Lerda, L. Magnea, R. Marotta and
R. Russo, ``String techniques for the calculation of renormalization
constants  in field theory'', Nucl.Phys.
{\bf B469} (1996) 235,  hep-th/9601143.}%
\nref\DiVecb {P. Di Vecchia, A. Lerda, L. Magnea, R. Marotta, ``Gauge theory
renormalizations from the open bosonic string'', Phys. Lett. {\bf B351}
(1995) 445, hep-th/9502156.} More recently, 
\nref\spain{E. Alvarez and C. G\'omez, ``Ultraviolet and Infrared
freedom from string amplitudes'', JHEP {\bf 9910} (1999) 018,
hep-th/9907205.}% 
\DiVec\DiVecb\spain\ calculated two,
three and four point gauge boson amplitudes in open bosonic string theory.
Using a suitable prescription for continuing the amplitudes off-shell, the
renormalization constants obtained from the zero slope limit of the string
theory amplitudes were observed to respect the background field gauge Ward
identities.  We will see below that the low energy limit of the string theory
amplitudes in our D-brane construction reproduce the field theory loop
amplitudes calculated in the background field gauge.

\ifig\oneloop{One loop contributions to the two-point function.}
{\epsfxsize6in\epsfbox{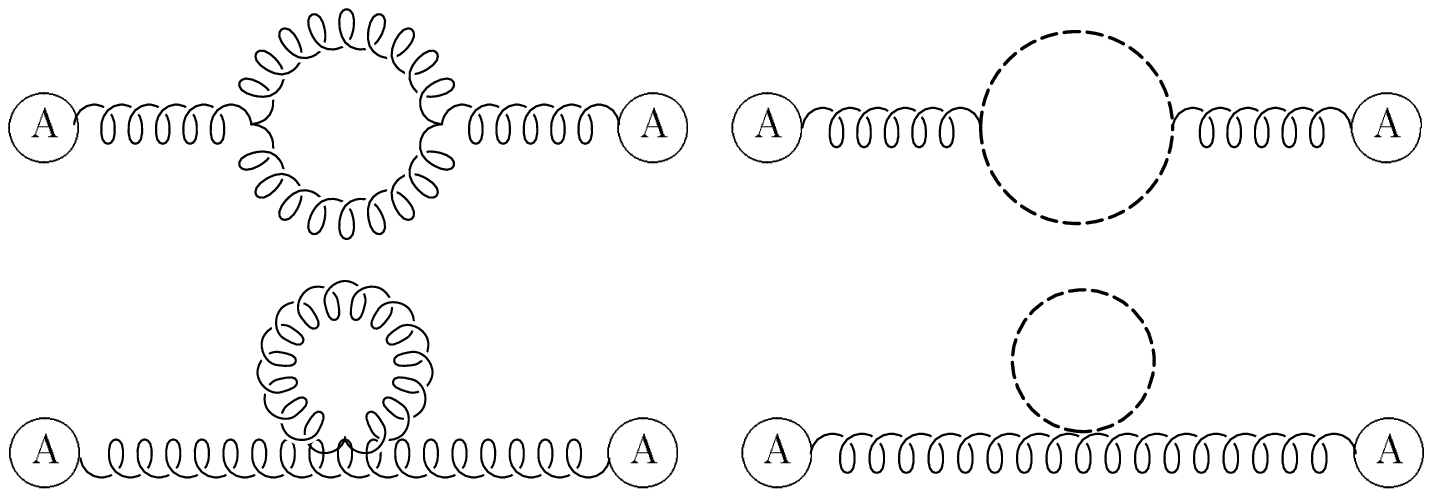}}

In \oneloop\ we show the one loop diagrams for the two-point function in U(1)
noncommutative gauge theory.  In ordinary gauge theory the tadpole diagrams
give vanishing contributions because \eqn\dr{ \int {d^dp \over p^2} = 0} in
dimensional regularization. However, in the noncommutative
 theory these graphs are nonvanishing and
must be included to obtain a gauge invariant answer.
The sum of the diagrams of Fig. 1 is
\eqn\oneloop{\eqalign{
\Pi_{\mu\nu}=\hskip-2pt
2 g^2 \hskip-2pt\int {d^dq \over (2 \pi)^d} {\rm sin}^2({{\tilde p} q\over 2})
\left[ {8 (p^2 G_{\mu \nu} - p_\mu p_\nu) \over q^2 (p+q)^2} +
 (d-2)\left( {(p+2 q)_\mu (p+2q)_\nu \over q^2 (p+q)^2} - {2 G_{\mu
 \nu}
\over q^2} \right) \right] ,}} where ${\tilde p}^\mu = \theta^{\mu
\nu}p_\nu$. Using the identity ${\rm sin}^2(x) ={1\over 2} (1- {\rm
cos}(2x))$ the field theory expression separates into two parts, one
independent of $\theta$ and one with a ${\rm cos}({\tilde p} q)$ in the
integrand. The term independent of $\theta$ corresponds to the planar
diagrams \filk ,
and gives an expression identical to ordinary Yang-Mills theory with the
usual group theory factor $f_{acd} f_{bcd} = N \delta_{ab}$ replaced by 2.
This piece is divergent in four dimensions and gives a $1/\epsilon$
pole.
 From this term one can extract the $\beta$-function of the
noncommutative theory \gkw\rjabbari\rruiz . The term with
the ${\rm cos}({\tilde p} q)$
corresponds to the nonplanar
 graphs and  is ultraviolet finite.

To compare with the string theory calculation in section $3$ it is
 best
 to combine the propagators
using Feynman parameters and then do the momentum integral via the method
 of
 Schwinger parameters.
The contribution of the planar graphs is
\eqn\planar{
\Pi_{\mu\nu}^P={i g^2 \mu^{4-d}\over (4 \pi)^{d/2}} (p^2 G_{\mu \nu} -
p_\mu p_\nu)
\int_0^{\infty} dt \int_0^1 dx \, t^{1-d/2} e^{-p^2 t x
 (1-x)}(8 -(d-2)(1-2x)^2).}
The contribution from the nonplanar graphs is \eqn\nonplanar{
\Pi_{\mu\nu}^{NP}\hskip-3pt =
-{i g^2 \over (4 \pi)^2} \int_0^{\infty} \hskip-3pt dt
\hskip-3pt\int_0^1 dx \,
 t^{-1} e^{-p^2 t x
(1-x)- {\tilde p}^2/4t} \hskip-3pt\left[ (p^2 G_{\mu \nu}\hskip-3pt -
p_\mu p_\nu) (8
-2(1-2x)^2)- \hskip-3pt{2 \over t^2} {\tilde p}_\mu  {\tilde p}_\nu\right]}
where since these diagrams are finite we have set $d=4$.

The planar diagrams are
ultraviolet divergent and to regulate
this divergence we take $d=
4- 2\epsilon$.
The result of evaluating the integral in \planar\ is
\eqn\plans{\eqalign{
\Pi_{\mu\nu}^P={i 2 g^2 \over (4 \pi)^2}(p^2 G_{\mu \nu} - p_\mu p_\nu )
\left(11-7\epsilon \over 3-2 \epsilon \right)
\left({-p^2 \over 4 \pi \mu^2}\right)^{-\epsilon}
 {\Gamma[\epsilon] \Gamma[1-\epsilon]^2 \over \Gamma[2-2\epsilon] } \cr
=  {i g^2 \over (4 \pi)^2} {22\over 3}(p^2 G_{\mu \nu} - p_\mu p_\nu )
\left({1\over \epsilon} - {\rm ln}\left(-p^2 \over \mu^2\right)
+...\right),}}
where $...$ is a constant.

The nonplanar diagrams evaluate to
\eqn\nonans{\eqalign{
\Pi_{\mu\nu}^{NP}&=-{i g^2 \over (4 \pi)^2} \int_0^1 dx
\left[ 2 (p^2 G_{\mu \nu} - p_\mu p_\nu )(8 -2(1-2x)^2) K_0 (p {\tilde p}
\sqrt{x(1-x)}) \right. \cr
 &\left. -16 {{\tilde p}_\mu  {\tilde p}_\nu \over {\tilde p}^4}
p^2 {\tilde p}^2 x(1-x) K_2(p {\tilde p} \sqrt{x(1-x)})  \right] \cr
&= -{i
g^2 \over (4 \pi)^2} \left[(p^2 G_{\mu \nu} - p_\mu p_\nu )\left( - {22\over
3}{\rm ln}\left( p^2 {\tilde p}^2\right)+...\right) - 32 {{\tilde p}_\mu
{\tilde p}_\nu \over {\tilde p}^4}+...\right] },}
where we have expanded in $p^2{\tilde p}^2$ to lowest order and kept
the most infrared singular terms.
 The nonplanar graphs give rise to ${\rm ln}(p^2
{\tilde p}^2)$, as first
 observed in  \mvs ,
as well as the correction to the photon polarization tensor of the form
${\tilde p}_\mu {\tilde p}_\nu /{\tilde p}^4$, observed in \haya\sunew . This
last term is interesting since it modifies the photon dispersion
relation.

We will see in the next section that the zero slope limit
 of the one loop string theory amplitude
exactly reproduces the field theory answer \planar\nonplanar .
 The Schwinger parameter $t$ in the field
theory calculation is proportional to the modulus of the string world sheet
while the
Feynman parameter $x$ is related to the separation of the
 vertex operators on the worldsheet.

\newsec{The String Theory Calculation}

In this section we reproduce the field theory results of section $2$ using
string theory. We will find a simple realization of the noncommutative U(1)
gauge theory using D3-branes in bosonic string theory and perform a one loop
string calculation which will yield, in the massless open string boundary of
moduli space, the results of the previous section.

A four dimensional  pure U(1) gauge theory can be realized by taking
the low energy limit
of the theory on a single D3-brane of bosonic string theory stuck at
an $R^{22}/Z_2$
orbifold singularity. This can be accomplished \mrd\ddg\ by
choosing the action
of the orbifold group
on the Chan-Paton factors to be represented by either of the two
one-dimensional
representations of $Z_2$. The projection equation projects out the
transverse scalars and
we are left with only a gauge field. The noncommutative version of
the pure U(1) gauge
theory can be obtained by applying a constant magnetic field along the D-brane
worldvolume.

The quantum effective action of this gauge theory is encoded in the string
theory effective action. An appropriate truncation of a string loop diagram
should provide, in the low energy limit, the field theory answer. We will
explicitly verify at the one loop level that the planar and non-planar
two-point amplitudes for the photon on the cylinder in a background magnetic
field reproduce the corresponding computations in the noncommutative gauge
theory in the background field gauge.

The correlation function of the photon field vertex operators on the disk
shows that the low energy  classical action on the brane is a noncommutative
gauge theory with the usual replacement of conventional products by
$\star$-products \seibwitt . Comparison
between the string theory calculation and the
field theory fixes the normalization of the photon vertex operator to be
\eqn\vertex{ V(z,k)=ig\int_{\partial \Sigma}ds\ e\cdot\partial_sXe^{ik\cdot
X},} where $g$ is the tree level Yang-Mills coupling constant, $e_{\mu}$ is
the polarization of the photon, $s$ is the coordinate on the worldsheet
boundary $\partial\Sigma$ and indices are contracted with the
$G^{\mu\nu}$ metric.

The worldsheet topology of the one-loop diagram is a cylinder, which we
represent in the complex z-plane as a rectangle of width $\pi$ and height
$2\pi i t$ -- where $0\leq t\leq \infty$ is the modulus of the cylinder --
and with the edges $y=0$ and $y=2\pi i t$ identified: \eqn\cylinder{\eqalign{
z&=x+iy\qquad 0\leq x\leq \pi\cr y&\simeq y+2\pi t\qquad 0\leq y\leq 2 \pi
t.}} Open string vertex operators must be inserted on the boundaries of the
cylinder at either $x=0$ or $x=\pi$, with the positions given by $w=iy$ and
$w=\pi+iy$ respectively.

The full two-point function is obtained by summing over the planar (two
vertex operators on the same boundary) and non-planar (each vertex operator
on a different boundary) diagrams. These diagrams are given by \eqn\ampli{
A=\int_{0}^{\infty}{dt\over 2t}\int_{0}^{2\pi t}dy_1 \int_{0}^{2\pi t}dy_2\
Z(t)<V(y_1,k_1)V(y_2,k_2)>\equiv e_1^\mu e_2^\nu\Pi_{\mu\nu},} where one
should keep in mind when computing $<\ldots>$ if the diagram is planar or
non-planar. The different terms in \ampli\ are easily understood. The $1/2t$
factor arises from explicitly gauge fixing the path integral and dividing by
the conformal Killing volume of the cylinder (which allows all vertex
operators to be unfixed). $Z(t)$ is the open string partition function of the
vacuum under consideration, which in our case is that of an open string
ending on  D3-brane of bosonic string theory stuck at an orbifold singularity
and in a background magnetic field. The correlator $<\ldots>$ is computed by
contracting the fields using the Green's function on the cylinder with
boundary conditions modified by the background magnetic field.

The open string partition function is a key ingredient in the measure of the
correlation function \ampli .  Worldsheet consistency conditions require
projecting the open string spectrum onto states invariant under the action of
the orbifold group. For our $Z_2$ orbifold this is reflected in the partition
function
\eqn\part{ Z(t)=\hbox{Tr}( {1+g\over 2}e^{-2tH_o})=Z_1(t)+Z_g(t),}
where $H_o$ is the open string Hamiltonian and $g$ is the $Z_2$ generator.
The $Z_2$ action on the endpoints of the string corresponding to a stuck
D3-brane just multiplies \part\ by unity
\nref\dg{D.E. Diaconescu and J. Gomis, ``Fractional Branes and
Boundary States in Orbifold Theories'', hep-th/9906242.}%
\mrd\ddg\dg .
 It is straightforward to show that
\eqn\unt{\eqalign{ Z_1(t)&=i\hbox{det}(g+2\pi \alpha^{\prime}F) {V_{p+1}\over
2}(8\pi^2\alpha^{\prime}t)^{-{p+1\over 2}}\eta(it)^{-24}\cr
Z_g(t)&=i\hbox{det}(g+2\pi \alpha^{\prime}F) 2^{{25-p\over 2}}{V_{p+1}\over
2}(8\pi^2\alpha^{\prime}t)^{-{p+1\over 2}}\vartheta_2(0,it)^{{p-25\over
2}}\eta(it)^{{27-3p\over 2}},}} where we have left explicit the
dimensionality of the brane (which will turn out to be useful when comparing
to the field theory results from section $2$). $F_{\mu\nu}$ are the
components of the background magnetic field.

In order to compute the correlation function \ampli\ we must solve for the
Green's function of the worldsheet scalars on the cylinder. The background
magnetic field along the brane does not modify the equations of motion of the
open strings ending on it, but it does change the boundary conditions on the
worldsheet fields. Worldsheet coordinates along the brane satisfy the
following boundary conditions\foot{The coordinates transverse to the brane
are projected out by the orbifold quotient.}: \eqn\bound{
g_{\mu\nu}\partial_nX^\nu+2\pi i
\alpha^{\prime}F_{\mu\nu}\partial_sX^\nu\Big|_{\partial\Sigma}=0.}  Here
$g_{\mu\nu}$ is the closed string metric (the metric that appears in the
string sigma model action). The operators $\partial_n$ and $\partial_s$ are
derivatives normal and tangential to the worldsheet boundaries
$\partial\Sigma$. The correlation function of vertex operators $<\ldots>$ on
a given worldsheet is computed from the propagators of the worldsheet fields,
which can be found  by solving the worldsheet wave equation while taking into
account the boundary conditions \bound . On the cylinder, the wave equation
to be solved is \eqn\wave{ {2\over \alpha^{\prime}}\partial_w\partial_{{\bar
w}}G^{\rho\sigma}(w,w^\prime)=-2\pi \delta^2(w-w^\prime)g^{\rho\sigma}+{1\over 2\pi t}g^{\rho\sigma}.} The last term,
which is proportional to the inverse area of the cylinder, is included in
order to satisfy Gauss' law and is compatible with the boundary conditions
\bound . The propagator we are interested in should solve \wave , satisfy
\bound\ at both boundaries of the cylinder, and respect the identification
$y\simeq y+2\pi t$ of the cylinder.
 The solution is given by
\eqn\prop{\eqalign{ {\cal G}&^{\mu\nu}(w,w^\prime)\equiv
<X^\mu(w)X^\nu(w^\prime)>= -\alpha^{\prime}\Bigg[ g^{\mu\nu}\Big(\log
\left|\vartheta_1({w-w^\prime\over 2\pi it}|{i\over t})\right|-\log
\left|\vartheta_1({w+ {\bar w^\prime}\over 2\pi it}|{i\over t})\right|\cr
-&{\hbox{Re}^2(w+{\bar w^\prime})+\hbox{Re}^2(w-w^\prime)\over 4\pi t} \Big)
+G^{\mu\nu}\log \left|\vartheta_1({w+{\bar w^\prime}\over 2\pi i t}|{i\over
t})\right|^2 +{\theta^{\mu\nu}\over 2\pi \alpha^{\prime}}\log
\left(-{\vartheta_1({w+{\bar w^\prime}\over 2\pi i t }|{i \over t})\over
\vartheta^*_1({w+ {\bar w^\prime}\over 2\pi i t }|{i \over
t})}\right)\Bigg],}} where $w$ and $w^\prime$ are points on the cylinder. 

The propagator on the cylinder has a similar structure to the propagator on
the disk.  Here $G^{\mu\nu}$ is the open string metric (the metric that
defines the dispersion relation for open string fields), and
$\theta^{\mu\nu}$ is the noncommutativity parameter which appears in the
definition of the $\star$-product. They are defined by
\eqn\parameters{\eqalign{ G^{\mu\nu}&=\left({1\over g+2\pi
\alpha^{\prime}F}g{1\over g-2\pi \alpha^{\prime}F}\right)^{\mu\nu}\cr
\theta^{\mu\nu}&=-(2\pi \alpha^{\prime})^2\left({1\over g+2\pi
\alpha^{\prime}F} F{1\over g-2\pi \alpha^{\prime}F}\right)^{\mu\nu}.}} The
noncommutative field theory is obtained by taking the limit
  $\alpha^\prime\sim\epsilon \rightarrow 0$ and $g\sim
\epsilon^{1/2}\rightarrow 0$ , with the magnetic field kept constant
\seibwitt .

For the two-point function of photon vertex operators we only need the
 propagator for
points on the boundaries. The correlator is given by \eqn\contract{\eqalign{
<&\partial_sX^\mu e^{ik\cdot X}(w)\partial_sX^\nu
e^{-ik\cdot
X}(w^\prime)>\cr
=&\left(\partial_{w}\partial_{w^{\prime}}{\cal
G}^{\mu\nu}+k_\rho k_\sigma \partial_{w}{\cal
G}^{\mu\rho} \partial_{w^{\prime}}{\cal
G}^{\sigma\nu}\right)e^{k_\rho k_\sigma({\cal
G}^{\rho\sigma}(w,w^\prime)-{1\over 2}{\cal
 G}_r^{\rho\sigma}(w,w)-{1\over 2}{\cal
 G}_r^{\rho\sigma}(w^\prime,w^\prime)
)}\cr
=&\left(-k_\rho
k_\sigma\partial_w{\cal G}^{\rho\sigma}\partial_{w^{\prime}}{\cal
G}^{\mu\nu}+k_\rho k_\sigma \partial_{w}{\cal G}^{\mu\rho}
\partial_{w^{\prime}}{\cal G}^{\nu\sigma}\right)e^{k_\rho k_\sigma({\cal
G}^{\rho\sigma}(w,w^\prime)-{1\over 2}{\cal
G}_r^{\rho\sigma}(w,w)-{1\over 2}{\cal
G}_r^{\rho\sigma}(w^\prime,w^\prime))}}}
 where we have
integrated the first term  by parts. ${\cal
G}_r^{mn}(w,w)$ is the renormalized propagator, which regulates the
divergences in the self-contractions by subtracting the short distance
behaviour of the propagator. The proper renormalized propagator for open
string vertex operators is given by
\eqn\renorm{ {\cal
G}_r^{\rho\sigma}(w,w^\prime)={\cal
G}^{\rho\sigma}(w,w^\prime)+\alpha^{\prime}G^{\rho\sigma}(\log|w-w^\prime|^2),}
where $w$ and $w^\prime$ are points on the same boundary.  We will denote the
combination in the exponent in \contract\ as
\eqn\expe{ {\widetilde{\cal
G}}^{\rho\sigma}(w,w^\prime)\equiv{\cal G}^{\rho\sigma}(w,w^\prime)-
{1\over 2}{\cal
G}_r^{\rho\sigma}(w,w)-{1\over 2}{\cal
G}_r^{\rho\sigma}(w^\prime,w^\prime)}
We will now
consider in turn the results for the planar and non-planar diagrams.

\subsec{Planar Two-point function}

${\widetilde{\cal
G}}^{\rho\sigma}(w,w^\prime)$
differs from ${\cal G}^{\rho\sigma}
(w,w^\prime)$ by a term that  is independent of the position of the vertex
operators. On the $x=0$ boundary it is given by\foot{Unlike Eq.(3.8)
above, this expression is not manifestly periodic. However, it is 
straightforward to rewrite this expression in a form in which
periodicity is manifest.}
\eqn\sameb{\eqalign{{\widetilde{\cal
G}}^{\rho\sigma} (w,w^\prime)&=-\alpha^\prime \left[G^{\rho\sigma}
 \log\left|2\pi{\vartheta_1({i(y-y^\prime)\over 2\pi},it)\over
\vartheta_1^\prime(0,it)}\right|^2-{(y-y^\prime)^2\over 2\pi
 t}\right]+{i\over 2}{\theta}^{\rho\sigma}\epsilon(y-y^\prime)\cr
&\equiv -\alpha^\prime G^{\rho\sigma }\Gamma(y-y^{\prime})+{i\over
2}{\theta}^{\rho\sigma}\epsilon(y-y^\prime),}}
where $w=y, w^\prime=y^\prime$ (or $w=\pi+iy, w^\prime=\pi+iy^\prime$),
$G^{\rho\sigma}$ is the open string metric and $\epsilon(x)$ is $1$
for $x>0$ and -1 for $x<0$. On the $x=\pi$ boundary the sign of the
term proportional to $\theta^{\rho\sigma}$ changes sign\foot{We thank
H. Dorn for correspondence on this point.}. 
Note that we have used a
$\vartheta$-function identity to rewrite the propagator in a form
conducive to taking the $t \rightarrow \infty$ limit.

Plugging this expression into the correlator \contract\ we see that it has
the familiar form of the vacuum polarization diagram of the photon
\eqn\express{ <\partial_sX^\mu e^{ik\cdot X}(w_1)\partial_sX^\nu e^{-ik\cdot
X}(w_2)>= -\alpha^{\prime 2}\left(k^2G^{\mu\nu}-k^\mu k^\nu
\right)(\partial_y \Gamma)^2e^{-\alpha^\prime k^2\Gamma},} where
$k^2=G^{\rho\sigma}k_\rho k_\sigma$ and $y=y_1-y_2$.

Combining all the terms in \ampli\ one is led to the
 following expression for the
planar two-point function \eqn\plan{
\Pi^P_{\mu\nu}=-g^2\int_{0}^{\infty}{dt\over
2t}\int_{0}^{2\pi t}dy_1 \int_{0}^{2\pi t}dy_2\ Z(t)\ \alpha^{\prime
2}\left(k^2 G_{\mu\nu}-k_\mu k_\nu\right)(\partial_y \Gamma)^2e^{-\alpha^\prime
 k^2\Gamma},}
where $Z(t)$ is given by \unt\ and $\Gamma$ by \sameb .

The task at hand is to identify in this string computation the noncommutative
field theory result. We have to examine \plan\ in the decoupling limit
specified in \seibwitt , which in particular requires taking the
$\alpha^\prime\rightarrow 0$ limit. In this limit we only get contributions
from  corners of string moduli space. We will now show that we obtain the
exact planar field theory answer from the boundary of moduli space of the
cylinder which is dominated by massless open strings\foot{Note that since we
are using bosonic string theory the field theory result is obtained only
after removing by hand the divergence caused by the open string tachyon.},
which comes from the $t\rightarrow \infty$ limit. We therefore need the large
$t$ expression of the integrand in \plan . The large $t$ expansions of $Z(t)$
and of $\partial_y\Gamma$ are given by \eqn\asympt{\eqalign{ Z(t)&\simeq
V_{p+1}(8\pi^2\alpha^{\prime}t)^{-{p+1\over 2}}\left(e^{2\pi t}+p-1+{\cal
O}(e^{-2\pi t})\right)\cr \partial_y\Gamma&\simeq 1-2x+2\left(e^{-2\pi
xt}-e^{2\pi xt}e^{-2\pi t}\right),}} where $x=y/2\pi t$. Plugging these
expressions into \plan\ and tossing out the contribution due to the tachyon,
one gets (with $d=p+1$ and $g\rightarrow g \mu^{4-d}$) \eqn\finalplanar{
\Pi_{\mu\nu}^P=i{g^2\mu^{4-d}\over (4\pi)^{d/2}}\left(k^2 G_{\mu\nu}-k_\mu
k_\nu\right)\int_0^\infty dt \int_0^1 dx\
t^{1-d/2}\left(8-(d-2)(1-2x)^2\right) e^{-k^2tx(1-x)},} which is precisely
the field theory answer \planar. To obtain this result we rescaled $t
\rightarrow t/\alpha^\prime$ and $y \rightarrow y/\alpha^\prime$.   In these
new variables (with $\alpha^\prime \rightarrow 0$) any finite value of $t$ is
in the extreme open string limit of the moduli space.    In particular the
excited open string corrections in \asympt\ become ${\cal O}(e^{-2\pi
t/\alpha^\prime})$ which vanish in the decoupling limit.  This will be
discussed further in Section $4$.

 \subsec{Non-Planar Two-point function}

The nonplanar propagator with $w=\pi+iy$ and $w^\prime=y^\prime$
is given by
 \eqn\nonp{\eqalign{
{\widetilde{\cal G}}^{\rho\sigma}= & -\alpha^\prime \Bigg[
G^{\rho\sigma}\left(
 \log\left|2\pi{\vartheta_2({i(y-y^\prime)\over 2\pi},it)\over
\vartheta_1^\prime(0,it)}\right|^2+{\pi \over 2t}-{(y-y^\prime)^2\over 2\pi
 t}\right)
+i{\theta^{\rho\sigma}\over 2\pi \alpha^\prime } {(y-y^\prime)\over t} \cr &
-g^{\rho\sigma} {\pi \over 2t} \Bigg].}} As in the planar diagram computation
we expand \nonp\ in the large $t$ region. The asymptotics of the term
proportional to the open string metric $G^{\rho\sigma}$ is the same as for
the planar diagram. The important differences are in the terms proportional
to the closed string metric $g^{\mu\nu}$ and the noncommutativity parameter
$\theta^{\mu\nu}$.  The term proportional to $\theta^{\mu\nu}$ does not
contribute to the exponential in \contract , but it plays a very important
role in the derivative terms.

Since this is a non-planar diagram
for an oriented string, there is an overall factor of $-1$ since the ends of
the string carry opposite charges.
 The final answer is
given by
\eqn\nopo{\eqalign{ \Pi_{\mu\nu}^{NP}\hskip-3pt&=-i{g^2\mu^{4-d}\over
(4\pi)^{d/2}}\int_0^\infty \hskip-3pt \hskip-3pt dt\ \int_0^1 \hskip-5pt dx
\, t^{1-d/2}e^{-k^2tx(1-x)-{\tilde k}^2/4t} \cr & \times \Big(\left( k^2
G_{\mu\nu}-k_\mu k_\nu\right)\left(8-(d-2)(1-2x)^2\right) - {2\over
t^2}{\tilde k^\mu}{\tilde k^\nu}\Big).}} This expression is identical to that
obtained from the nonplanar field theory graphs \nonplanar.

\newsec{Discussion}

In the previous sections we have seen how the annulus (cylinder)
amplitude of string theory in a background $F$ field reproduces,
in the decoupling limit of \seibwitt,  the planar and nonplanar
results of noncommutative gauge theory.   Let us discuss this in
more detail. Schematically the two-point function on the annulus
is given in the
open string channel by
 \eqn\annulus{A \sim \int_0^{\infty}  {dt~ t} \sum_I\exp{(-\Delta_I t)}.}
The index $I$ labels all open string states and $\Delta_I$, basically the
$L_0$ eigenvalue, is the mass squared of state $I$ plus momentum dependence.
In the decoupling limit of \seibwitt\ $\alpha^\prime \rightarrow 0$ and hence
the string mass scale is sent to infinity. The oscillator contribution to
$\Delta_I$ is unaffected by $F$ and hence is just $N_I/\alpha^\prime$ where
$N_I$ is the total oscillator occupation number. So in the $\alpha^\prime
\rightarrow 0$ limit all the excited open string states become much heavier
than the massless one and hence should decouple from  the dynamics.  The
vanishing of the exponential corrections in \finalplanar\ and \nopo\
illustrates this. This is similar to other decoupling limits such as those
which show that field theories arise from branes separated by short
distances. As pointed out in \mvs\ and further discussed in \vs\sunew there
are peculiar singularities indicating IR/UV mixing in the nonplanar
noncommutative gauge theory results.   These results have been interpreted to
mean \mvs\vs\ that some closed string residue remains in the field theory,
even in the  decoupling limit.  So it is important to examine if and how
decoupling is breaking down here.

Generally,  the only way decoupling can fail is for the interactions of the
decoupled theory to have bad high energy behavior. If loops of massless open
string states are  UV divergent, then massive open string states will
generally be excited,  violating decoupling.   If the decoupled field theory
is divergent but renormalizable then there will be a mild violation of
decoupling, but all the effects of the massive string states can be absorbed
in a  few ``renormalized'' couplings.   For instance, $g^2 \ln({-p^2 /
{\mu^2}})$ in \plans\ becomes, in string theory, $g^2(\ln(-p^2 \alpha^\prime)
+ {\cal O} (1))$.

Now let us examine decoupling in the nonplanar diagrams that
display mysterious UV/IR singularities.  We can write a caricature
of the string amplitude \nopo\ by suppressing the $x$ integral, all
numerical factors, and adding back in the effect of the first
excited  open string state. \eqn\caric{ A_{NP}\sim  \int dt~
t^{1-d/2}e^{-k^2t-{\tilde k}^2/4t}~(1+e^{-2\pi t/\alpha^\prime}+
\ldots).} Note that the ${\tilde k}^2/4t$ term in the exponential
renders the UV region of the modular integration ($t \rightarrow
0$) completely finite for any nonzero ${\tilde k}^2$. This term is
present in the field theoretic amplitude \nonplanar\ and represents
the effect of the rapidly oscillating phases in the noncommutative
gauge theory interaction vertices.  These phases are enough to
render the nonplanar amplitude finite for any nonzero ${\tilde
k}^2$.

The smallest important value of $t$ in \caric\ is roughly $t \sim
{\tilde k}^2$.   The contribution of the first excited state is
then
 $\sim e^{-2\pi{\tilde k}^2 /\alpha^\prime}.$  In the decoupling
 limit $\alpha^\prime \rightarrow 0$ and this contribution
 vanishes for any nonzero ${\tilde k}^2$.   The decoupled field
 theory amplitude is UV finite so decoupling cannot fail.

 To further investigate this question let us keep $\alpha^\prime$
 finite.   There are then two regimes to consider.   If
${\tilde k}^2 \gg \alpha^\prime$ then the excited open string state
contribution is negligible and the decoupled field theory result is accurate.
If ${\tilde k}^2 \ll \alpha^\prime$, however, the small $t$ region of the
integral may be important. This depends on whether the field theory graph
without phase factors is UV divergent.   It will be, for instance, if the
space-time dimension $d$ is large enough. If there is a small $t$ UV
divergence then all the excited open string states will become important.  In
this case the correct way to analyze the situation is to use channel duality
and rewrite the amplitude in terms of closed string states.  At small $t$
only the lightest closed string states will contribute, giving a massless
propagator $1/k^2$ (assuming we drop the closed string tachyon).

The region where the closed string description is valid becomes smaller and
smaller as we approach the decoupling limit $\alpha^\prime \rightarrow 0$. In
this limit the region of validity shrinks to a set of measure zero.   For all
finite ${\tilde k}^2$ the decoupled field theory describing only the lightest
open string mode is exact.  So the complete structure of the mysterious IR/UV
singularities is contained in the open string description.

Of course one can formally represent the behavior of the lightest
open string in the dual closed string channel.  But this requires
a sum over closed string states of arbitrarily high mass and
does not seem very transparent.  This is the usual situation in
dualities.  A regime that has a simple description in one set of
variables typically has a complicated description in the dual
variables.

To illustrate this point consider \caric\ for general $d$.  The log
divergence in $d=4$ becomes a $1/{\tilde k}^{d-4}$ divergence.
The massless closed strings will produce a $1/k^2$ behavior for
any $d$.  To produce the open string behavior will require a sum
over all the closed string states.

There is at least one situation where  decoupled field theory results can be
reproduced from the lightest closed string state\foot{This observation is due
to Lenny Susskind.}. This is the case \nref\dougli{M.R. Douglas and M. Li,
"D-Brane Realization Of
 N=2 Superyang-Mills Theory In Four-Dimensions," hep-th/9604041.}
\nref\dkps{M.R. Douglas, D. Kabat, P. Pouliot and S. H. Shenker,
"D-Branes And Short Distances In String Theory," Nucl. Phys.
{\bf B485} 85 (1997).} \nref\backir{C. Bachas and E. Kiritsis,
"F(4) Terms In N=4 String Vacua,"  Nucl. Phys. Proc.Suppl.
{\bf 55B} 194(1997).}%%
where a nonrenormalization theorem exists \dougli\dkps\backir. The
contribution of excited open string states vanishes, typically because they
are in long multiplets of an extended supersymmetry algebra.   The exact
amplitude is given by the lightest open string state, and so this must also
agree with the closed string answer.   This mechanism seems to require lots
of supersymmetry, and usually applies only to special amplitudes. So it seems
questionable whether it will be helpful in giving a general explanation for
these mysterious singularities.

\medskip
\medskip
\medskip
\medskip

\centerline{\it Note Added}

In the past week the papers
\nref\ad{O. Andreev and H. Dorn, "Diagrams of Noncommutative $\Phi^3$ Theory
from String Theory", hep-th/0003113.}
\nref\kilee{Y. Kiem and S. Lee, ``UV/IR Mixing in Noncommutative Field
Theory via Open String Loops'', hep-th/0003145.}%
\nref\bcr{A. Bilal, C.-S. Chu and R. Russo, `` String Theory and
Noncommutative Field Theories at One Loop'', hep-th/0003180.}%
\ad\kilee\bcr\ appeared on the archive. They overlap substantially with ours.

\centerline{\bf Acknowledgments}

We would like to thank Allan Adams, Sergei Gukov, John McGreevy, Hirosi Ooguri,
Maxim
Perelstein, John Schwarz, Lenny Susskind,  Nick Toumbas  and Edward Witten
for discussions. We
would like to thank Iain Stewart for help making the figures. J.G. and T.M.
are supported in part by the DOE under grant no. DE-FG03-92-ER
40701. M.R. is supported by the Caltech Discovery Fund under grant
no. RFBR 98-02-16575 and DE-FG-05-ER 40219.
 M.K. is supported by an NSF graduate fellowship. S.S. is
partially supported by NSF grant 9870115.

\nref\acny{A. Abouelsaood, C.G. Callan, C.R. Nappi and S.A. Yost, "Open
Strings in Background Gauge Fields", Nucl. Phys. {\bf B280} (1987) 599-624.}

\listrefs

\end